\documentclass{PoS}
\usepackage{subcaption}

\title{Measurement of the Diffuse Muon Neutrino Flux using Starting Track Events in IceCube}

\author{
The IceCube Collaboration\footnote{For collaboration list, see PoS(ICRC2019) 1177.}\\
{\itshape \href{http://icecube.wisc.edu/collaboration/authors/icrc19_icecube}{http://icecube.wisc.edu/collaboration/authors/icrc19\_icecube}}\\
E-mail: \email{msilva@icecube.wisc.edu, smancina@icecube.wisc.edu}
}

\abstract{

IceCube measures the diffuse neutrino flux across several neutrino flavors and energy ranges. ESTES focuses on the measurement of the diffuse neutrino flux using high purity astrophysical muon neutrinos with energies above 1 TeV. We use the Enhanced Starting Track Event Selection dataset which selects for tracks starting within the IceCube fiducial volume. We employ a machine learning algorithm to help differentiate between tracks from atmospheric and astrophysical neutrinos. This produces a high purity diffuse neutrino sample that can provide valuable insight into the properties of the atmospheric and astrophysical diffuse neutrino spectrum. Using simulated neutrinos, we show the sensitivity for this measurement of the diffuse neutrino flux with ESTES. \\

\vspace{4mm}
{\bfseries Corresponding authors:}
\speaker{Manuel Silva}$^{1}$, Sarah Mancina$^{1}$\\
{$^{1}$ \itshape Dept. of Physics and Wisconsin IceCube Particle Astrophysics Center, University of Wisconsin, Madison, WI 53706, USA}

}

\FullConference{36th International Cosmic Ray Conference -ICRC2019-\\
		July 24th - August 1st, 2019\\
		Madison, WI, U.S.A.}

\ShortTitle{ESTES Diffuse - 2019}

\begin{document}


\section{Introduction}\label{sec:info}
The Enhanced Starting Track Event Selection (ESTES) is an event selection using IceCube data currently under development. The primary goal is simply to obtain a high purity diffuse muon neutrino sample for neutrino energies between 1-100 TeV. However, due to the significant presence of atmospheric muons in this energy range, one must employ sophisticated veto techniques. Ongoing work to increase the neutrino purity is shown in this proceeding. \\
Once a high purity sample of neutrinos is obtained, many interesting physics topics can be explored. One possibility is to add this event selection to the real time alerts IceCube sends to the multi-messenger community \cite{ESTESv1:2019icrc_ESTESv1}. ESTES can also be used to measure the astrophysical diffuse neutrino flux, including (but not limited to) the overall normalization and spectral index. The large presence of atmospheric neutrinos suggests strong constraints can be placed on various atmospheric flux models. Estimates of the expected astrophysical neutrino rates for ESTES are shown in
this presentation. 


\section{Simulation}\label{sec:sim}

\subsection{Monte Carlo Samples}\label{sec:MC}
Atmospheric muons are generated using the CORSIKA package and include events with multiple muons. The livetime of this dataset corresponds to 1/10th of a year of data taking and consists of 100 million events. Additional single muon events are generated using a parametrized version of the muon flux at depth in ice. The livetime of this dataset corresponds to over 10 years and consists of over 3 billion events. When computing rates, the assumed cosmic-ray flux is from Gaisser H4a \cite{h3a_flux}.\\
The astrophysical neutrinos are injected near the IceCube surface and propagated to the IceCube volume. The neutrinos are forced to interact within IceCube with the probability of interaction taken into account as a scaling factor when deriving the rates. The astrophysical neutrinos rate of observation is now derived assuming a single power law flux with a predefined overall normalization and spectral index.

\subsection{Reconstructions}\label{sec:Reco}
The SplineMPE track reconstruction \cite{angular_res} calculates the muon track direction using the number of photo electrons deposited in each optical module along with timing information. This reconstruction has been shown to yield median angular resolution lower than 1$^\circ$ for a wide range of energies, including our energies of interest. We then estimated the energy losses along the SplineMPE muon track using the Millipede algorithm \cite{energy_res}. Millipede splits the total track into 10m long track segments and estimates the energy lost per segment. Since we are interested in starting tracks only, some energy is not measured when the track exits the detector, therefore these estimates serve only as a lower limit on the muon energy. At relevent energies, the total energy deposit can be reconstructed with resolution better than 10$\%$.


\section{Event Selection}\label{sec:ESTES}
\subsection{Previous Work}\label{sec:ESTES_startingtrack}
\begin{figure}[h!]
\centering 
    \includegraphics[width=0.85\linewidth]{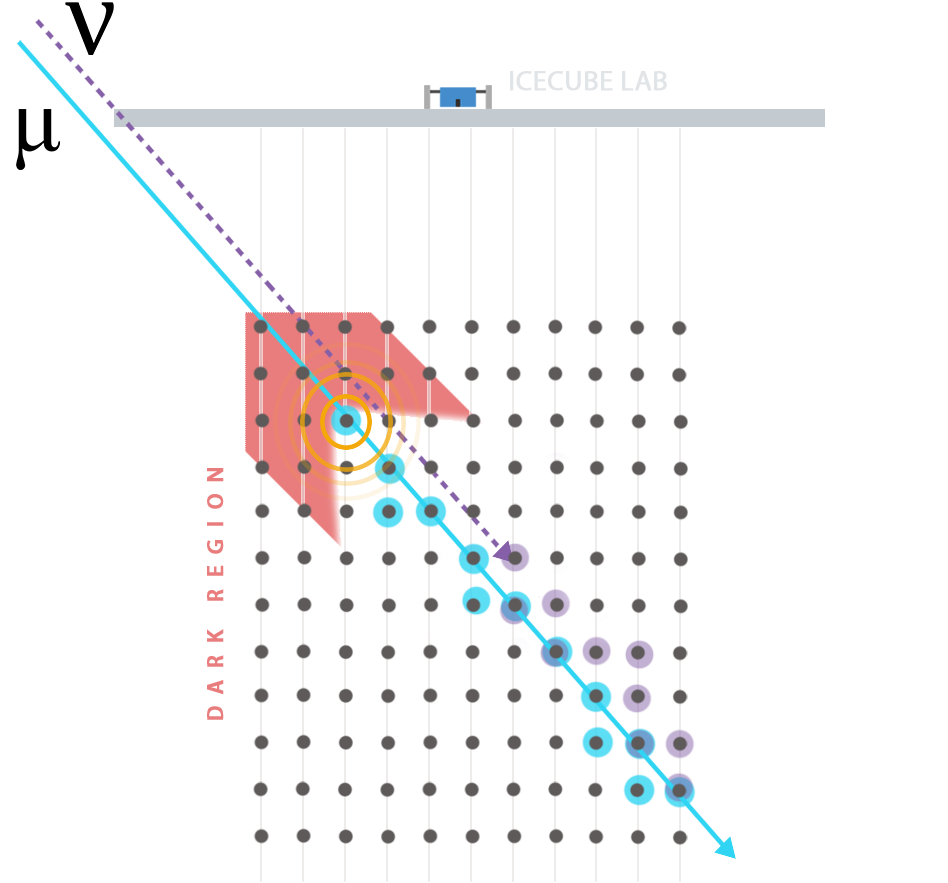}
    \caption{The dark region is shown in red and represents the region where one would expect to see energy losses given a muon that is continuously emitting electromagnetic radiation. This "dark region" is geometrically constructed using the Cherenkov cone produced from the location where the first PE is observed, eg the vertex. }
\label{fig:STV}
\end{figure}

ESTES can be summarized in 2 steps that each aim to heavily reduce atmospheric muon rates. The first step employs a starting track veto (STV) used to reduce the number of muon tracks in the data sample. Figure \ref{fig:STV} shows how this STV is executed for a particular muon track event. The STV works by computing a probability of the IceCube DOM to observe PEs given a muon track hypothesis in the dark region. A cut is then applied on this probability which reduces the expected atmospheric muon rates from over 100 million down to 10 thousand per year. To learn more details pertaining to this veto please see \cite{Kyle:ESTES} \& \cite{Mancina:ESTES}. A boosted decision tree (BDT) is then used which reduces the expected muon rate from about 10 thousand per year to 1 per year. Much effort was put into this BDT to maximize the number of starting track neutrinos in our final data sample.

\subsection{Boosted Decision Tree}\label{sec:ESTES_BDT}
We use the XGBoost BDT (eXtreme Gradient Boosting \cite{Chen:2016:XST:2939672.2939785})  to help filter atmospheric muons. The XGB algorithm was chosen due to resistance to over-training, speed, and high accuracy in correctly classifying neutrinos and muons.\\

\begin{table}[h!]
\footnotesize
\centering 
\begin{tabular}{|p{2.5cm}|p{2.5cm}|p{2.5cm}|p{2.5cm}|p{2.5cm}|}
\hline
Z of entry position & Closest distance to from edge, from first energy loss & Fraction of energy in first loss & Distance to edge along track & p$_{miss}$ from segmented track \\ \hline
Number of energy losses & Total energy of losses & Fraction of direct hits &  Length of total losses & Reco. Zenith Angle \\ \hline
Fraction of charge on detector edge & Number of fits in step 1 & Reco. Speed of track & p$_{miss}$ from infinite track & $\Delta \Theta$ of different recos. \\ \hline 
\end{tabular}
\caption{The 15 variables used to train the BDT and classify individual events.}
\label{tab:BDT_variables}
\end{table}

\begin{figure}[h!]
  \centering
  \begin{subfigure}[b]{0.49\linewidth}
    \includegraphics[width=\linewidth]{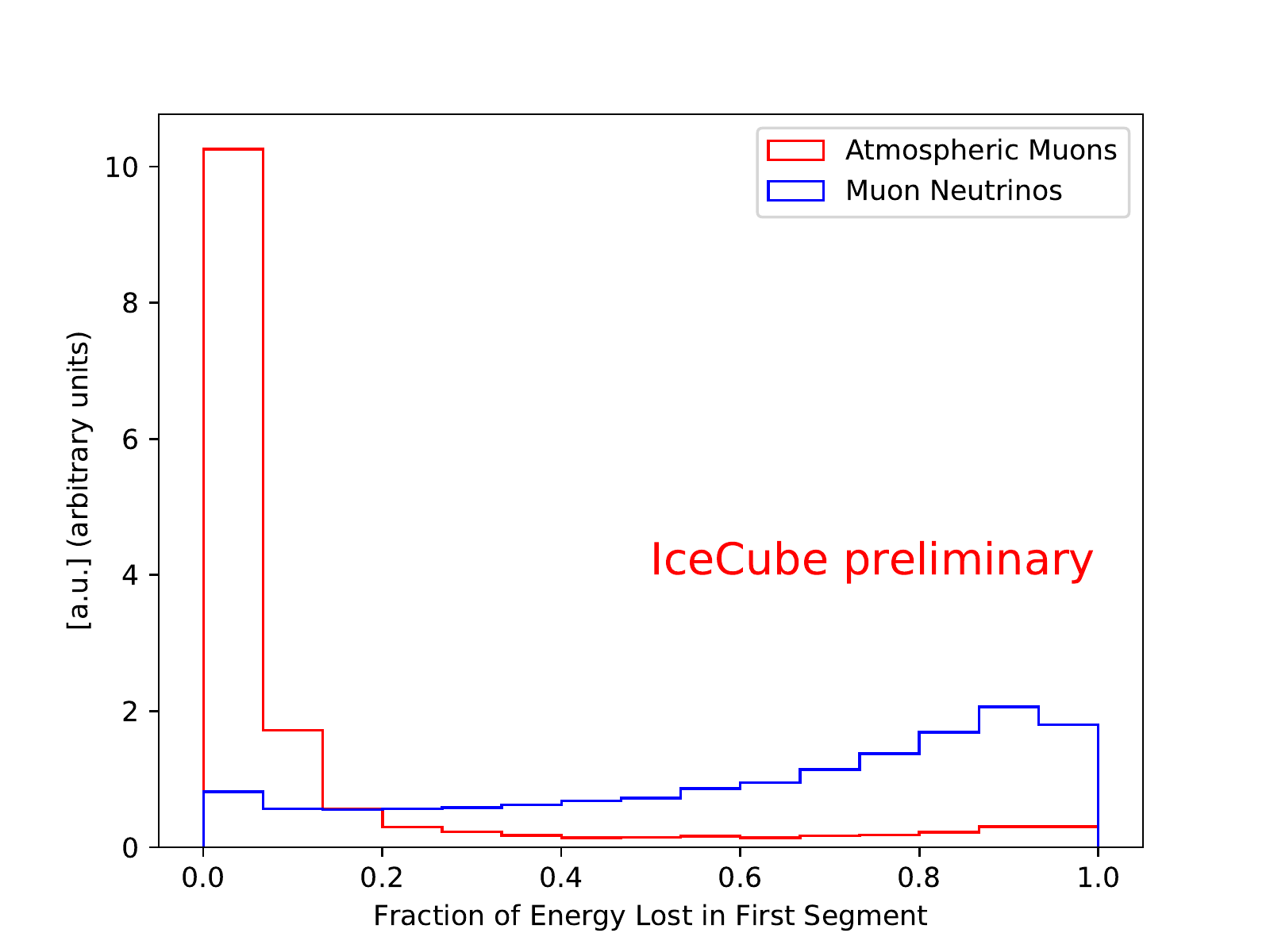}
  \end{subfigure}
  \begin{subfigure}[b]{0.49\linewidth}
    \includegraphics[width=\linewidth]{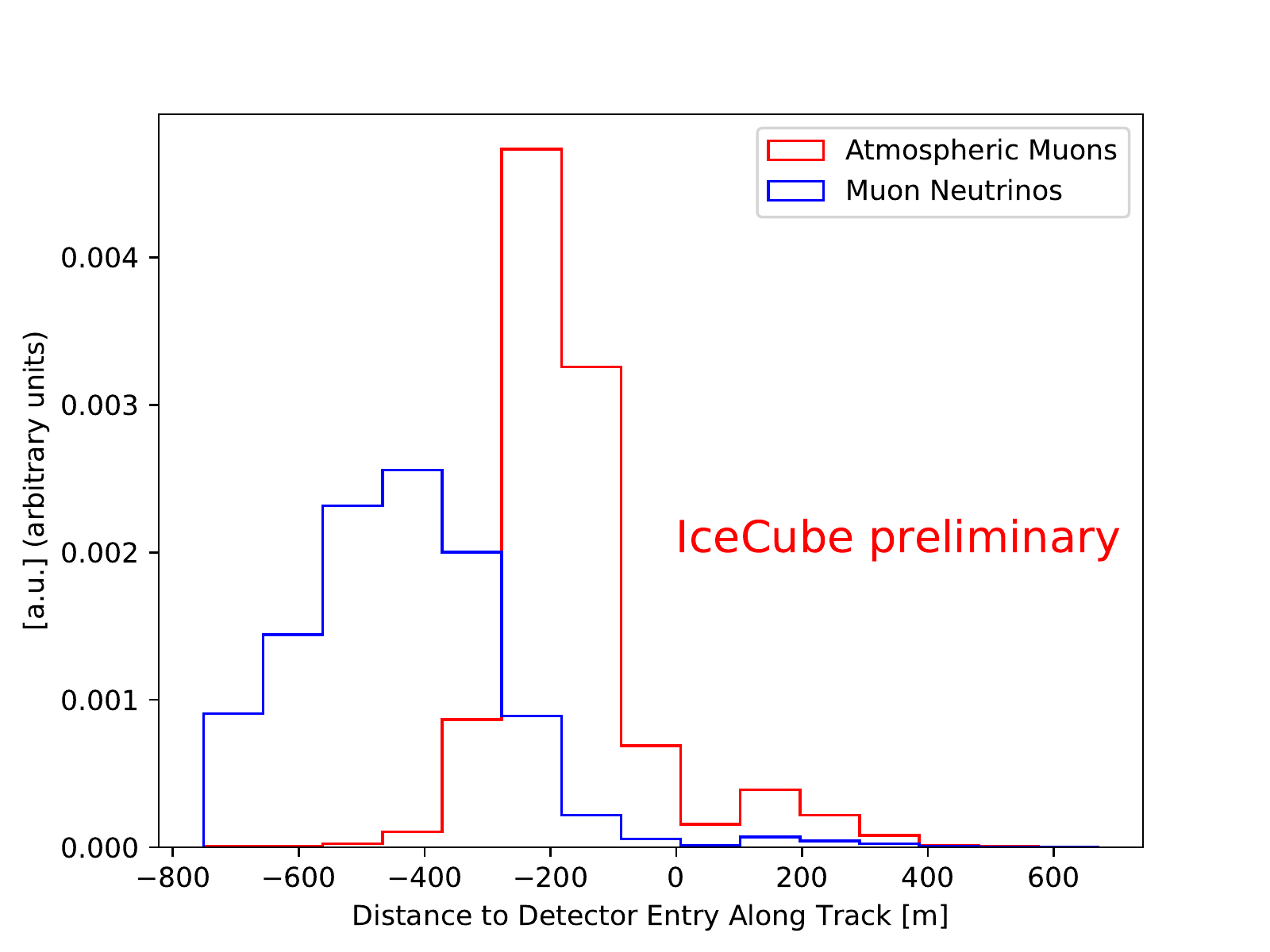}
  \end{subfigure}
  \caption{Most important inputs for training the BDT. No weights are applied to training therefore the distributions are exactly as seen by the BDT.}
  \label{fig:BDTinputs}
\end{figure}

\begin{figure}[h!]
\centering 
    \includegraphics[width=0.85\linewidth]{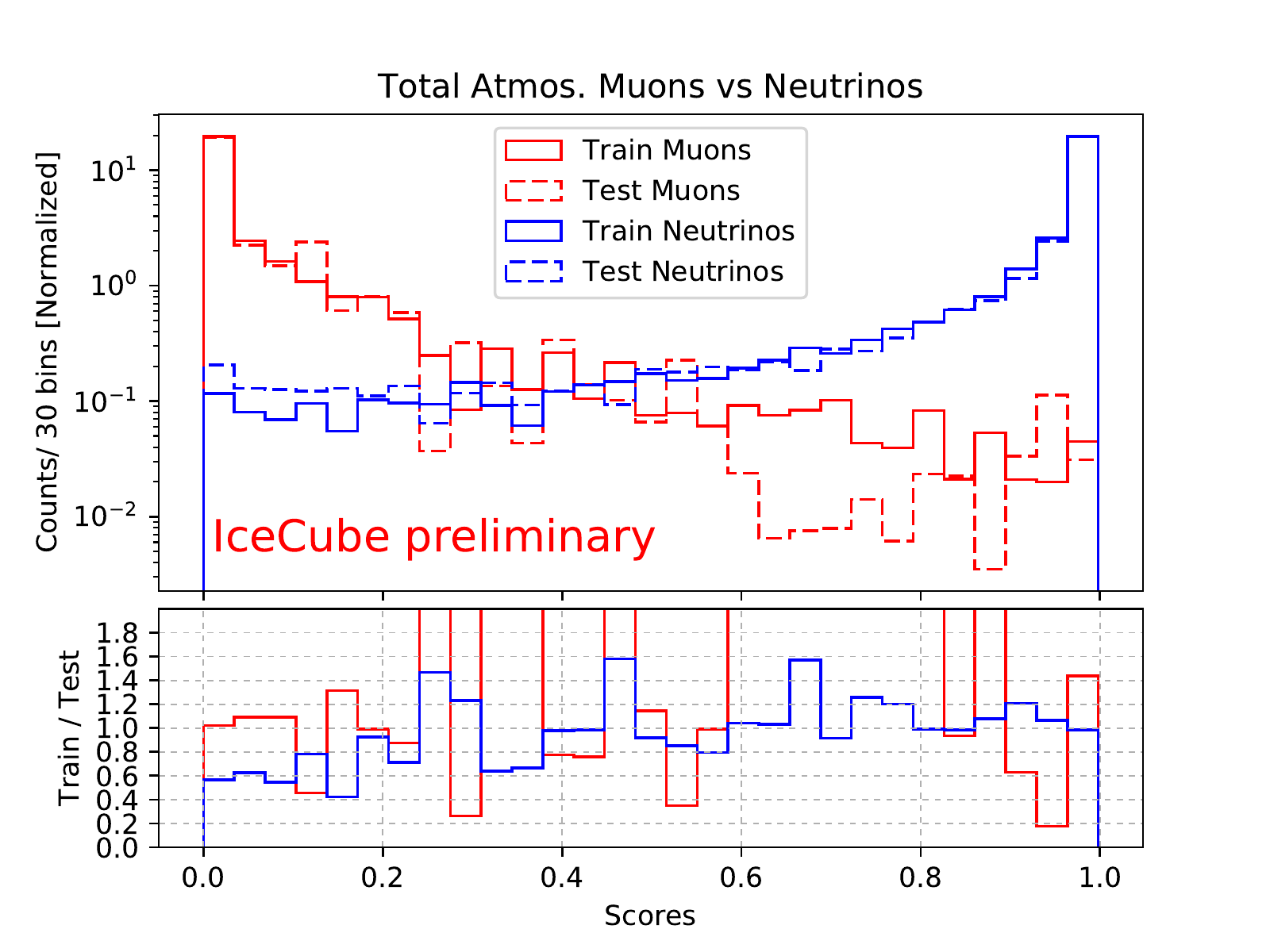}
    \caption{Probabilities computed using the trained BDT for both muons and neutrinos. The muons event rates are computed assuming the GaisserH4a cosmic ray flux, the neutrinos are weighted to the MESE flux.}
\label{fig:BDToutputs}
\end{figure}

The BDT was trained using the reconstructed parameters from section \ref{sec:Reco} as input. For a complete list of the input variables see table \ref{tab:BDT_variables}. After the training is completed, a ranking of the 15 input parameter's contribution to the accuracy is computed. The most important variable for training is the energy lost in the first segment divided by the total energy loss along the track, fraction of energy lost in first segment. The second most important input to the BDT is the expected distance from the reconstructed vertex to the estimated point of entry to IceCube. Figure \ref{fig:BDTinputs} shows these variables for both muons and muon neutrinos. After the BDT is trained, a BDT classification score is produced for each individual event. For the purposes of training a stable BDT, 30\% of the total Monte Carlo is used for testing purposes only and is exclusive of the training dataset. If the BDT is over-fit to the training Monte Carlo sample, the test sample would be in clear disagreement with the training sample. A two-sided Kolmogorov Smirnov test is used to test the probability that the train and test samples come from the same distribution. A p-value of 0.30 is obtained for both muons and neutrinos which is satisfactory to conclude that the BDT model is not over-fit. The distribution of BDT classification scores are shown in figure \ref{fig:BDToutputs} along with a ratio of the test to train datasets.

\begin{figure}[h!]
\centering 
    \includegraphics[width=0.85\linewidth]{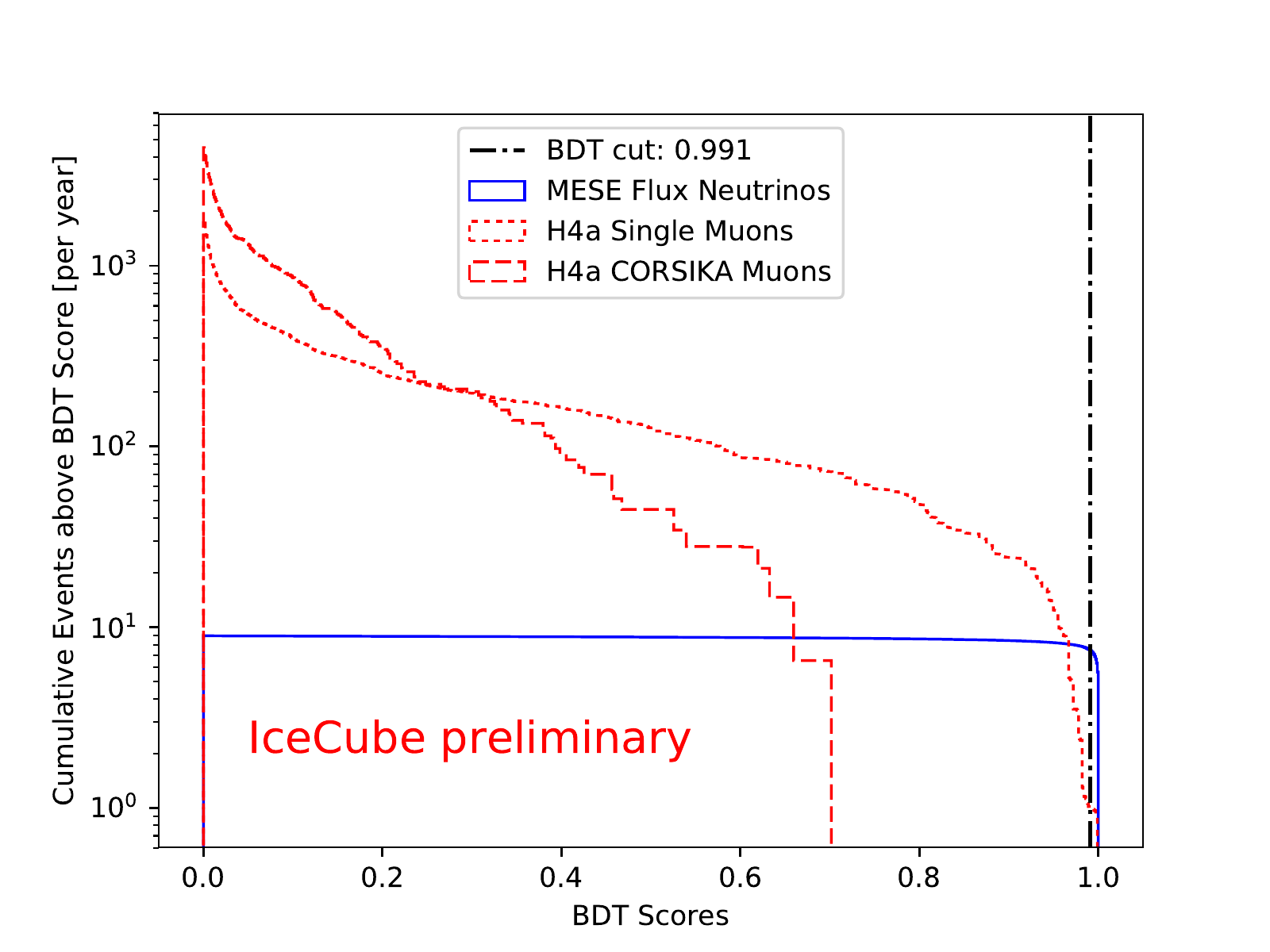}
    \caption{The weighted number of events integrated over the BDT score. The first bin represents the total number of expected muons and neutrinos entering the BDT. The optimized cut on BDT score is 0.991 and corresponds to an 80$\%$ efficiency at accepting neutrinos.}
\label{fig:BDT_cuts}
\end{figure}

Once the BDT is trained and optimized, a cut of 1 muon per year is chosen which corresponds to a BDT score of 0.991. All events above this cut will enter the final event selection. Figure \ref{fig:BDT_cuts} shows the cumulative number of muons and astrophysical neutrinos entering the final dataset given a particular BDT cut.

\section{Diffuse Neutrino Flux Sensitivity}\label{sec:Diffuse}
For the purpose of a sensitivity study, three IceCube fluxes are injected. These are provided as a differential flux in units 

\begin{equation}
\frac{\mathrm{d}\Phi}{\mathrm{dE}} = \Phi_{0} \times 10^{-18} \times (\frac{\mathrm{E}_{\nu}}{10^{5} \mathrm{GeV}})^{-\gamma} (\frac{1}{\mathrm{GeV\cdot cm^{2}\cdot sr\cdot s}})
\end{equation}

, where the two parameters of interest in the fit are $\Phi_{0}$ and $\gamma$. 
A summary of the measured astrophysical neutrino fluxes is presented in table \ref{tab:fluxes}. The injected fluxes are from the 3 different IceCube publications complimentary to each other due to event topology and neutrino energies. ESTES is similar to MESE \cite{MESE:2year} for the neutrino energies of interest, above 1 TeV, however MESE is designed to focus on cascade like events. HESE \cite{Austin:HESE} is similar to ESTES due to low background contamination in the final event selection. ESTES is most similar to the NuMu analysis \cite{icrc:numu} such that we are searching for muon neutrino events, however the NuMu measurement only uses northern sky tracks whereas we focus on southern sky numu tracks. The inelasticity result \cite{inelas} is similar to ESTES in that the focus is placed on starting track events, however this measurement is dominated bu northern sky events whereas ESTES will focus on high purity southern sky events. The various predicted rates (in units per year) are shown in table \ref{tab:fluxes} for astrophysical neutrinos. 

\begin{table}[h!]
\centering 
\begin{tabular}{l|c|c|r}
\hline
Injected Flux & $\Phi_{0}$ & $\gamma$ & N$_{exp.}$ (per year) \\ \hline \hline 
MESE\cite{MESE:2year} & 2.06 & 2.46 & 7.43 $\pm0.03$ \\ \hline
HESE\cite{Austin:HESE} & 2.15 & 2.89 & 16.07 $\pm0.08$ \\ \hline
NuMu\cite{icrc:numu} & 1.44 & 2.28 & 4.15 $\pm0.02$ \\ \hline
Inelas.\cite{inelas} & 2.04 & 2.43 & 7.43 $\pm0.04$ \\ \hline
\end{tabular}
\caption{Table of various IceCube measured flux central values. N$_{exp.}$ (per year) is the total number of expected astrophysical neutrinos after applying ESTES to the Monte Carlo, errors are derived from the MC directly. This is the rate at which we expect to observe astrophysical neutrinos assuming various diffuse neutrino spectra. Note that this result is statistical only.}
\label{tab:fluxes}
\end{table}


\section{Conclusion and Future Work}\label{sec:Conclusion}
The last step in the ESTES event selection was shown to yield negligible amounts of background contamination from atmospheric muons with an efficiency of over 80$\%$ to keep neutrinos with starting tracks. Various diffuse neutrino spectra were injected and yield comparable results.
\\
At this point, there are many checks undergoing to validate the overall event selection and prepare for data. The eventual goal is to measure the diffuse astrophysical neutrino flux using all IC-86 IceCube data which, up to ICRC 2019, corresponds to 8 years of data. Once the data is checked for compatibility with the Monte Carlo, a fit to the astrophysical diffuse neutrino spectrum will be performed. This will be followed by a release of a new real-time event stream to the multi-messenger community.


\bibliographystyle{ICRC}
\bibliography{references}

\end{document}